\documentclass[aps,pre,showpacs,twocolumn,groupedaddress]{revtex4-1}

\usepackage{graphicx}
\usepackage{amsbsy}
\usepackage{color}
\usepackage{hyperref}
\usepackage{breakurl}

\begin{document}
\title{Dynamics of a Semiflexible Polymer or Polymer Ring in Shear Flow}
\author{Philipp S. Lang$^{1,2}$}
\author{Benedikt Obermayer$^{1}$}
\altaffiliation{Current address: Max-Delbr\"uck Center for Molecular Medicine, Robert-R\"ossle-Str. 10, D-13092 Berlin, Germany}
\author{Erwin Frey$^{1,2}$}
\email[]{frey@lmu.de}
\affiliation{$^1$Arnold-Sommerfeld-Center for Theoretical Physics and Center for NanoScience, Department of Physics, Ludwig-Maximilians-Universit\"at M\"unchen, Theresienstrasse 37, D-80333 Munich, Germany}
\affiliation{$^2$Nanosystems Initiative Munich (NIM), Ludwig-Maximilians-Universit\"at M\"unchen, Schellingstra\ss e 4, D-80333 Munich, Germany}
\date{\today}

\begin{abstract}
Polymers exposed to shear flow exhibit a remarkably rich tumbling dynamics. While rigid rods rotate on Jeffery orbits, a flexible polymer stretches and coils up during tumbling. Theoretical results show that in both of these asymptotic regimes the corresponding tumbling frequency $f_c$ in a linear shear flow of strength $\gamma$ scales as a power law $W\!i^{2/3}$ in the Weissenberg number $W\!i = \gamma \tau$, where $\tau$ is a characteristic time of the polymer's relaxational dynamics. 
For a flexible polymer these theoretical results are well confirmed by a large body of experimental single molecule studies. However, for the intermediate semiflexible regime, especially relevant for cytoskeletal filaments like F-actin and microtubules, the situation is less clear. While recent experiments on single F-actin filaments are still interpreted within the classical  $W\!i^{2/3}$ scaling law, theoretical results indicated deviations from it. 
Here we perform extensive Brownian dynamics simulations to explore the tumbling dynamics of semiflexible polymers over a broad range of shear strength and the polymer's persistence length $l_p$. We find that the Weissenberg number alone does not suffice to fully characterize the tumbling dynamics, and the classical scaling law breaks down. 
Instead, both the polymer's stiffness and the shear rate are relevant control parameters.  Based on our Brownian dynamics simulations we postulate that in the parameter range most relevant for cytoskeletal filaments there is a distinct scaling behavior with $f_c \tau^* = W\!i^{3/4} {\hat f}_c (x)$ with $W\!i=\gamma \tau^*$ and the scaling variable $x = (l_p/L)(W\!i)^{-1/3}$; 
here $\tau^*$ is the time the polymer's center of mass requires to diffuse its own contour length $L$. Comparing these results with experimental data on F-actin we find that the $W\!i^{3/4}$ scaling law agrees quantitatively significantly better with the data than the classical $W\!i^{2/3}$ law. Finally, we extend our results to single ring polymers in shear flow, and find similar results as for linear 
polymers with slightly different power laws.
\end{abstract}

\maketitle

\section{Introduction}

Conformations as well as dynamics of biopolymers are nowadays well accessible through single molecule studies. Biopolymers with different degrees of flexibility like DNA \cite{BustamanteNature2003, PerkinsChu1994Science, LeDucWirtz1999Nature} or cytoskeletal filaments like F-actin \cite{leGoffFrey2002prl, ott1993measurement} and microtubules \cite{pampaloni2006thermal, taute2008microtubule, caspi1998semiflexible, janson2004bending, brangwynne2007bending} have been studied extensively.  
This has, in combination with theoretical efforts, lead to important insights into the statistics of their conformations in thermal equilibrium \cite{shimada1984ring, wilhelm1996radial, Everaers2010EuropPhysJE}. Using the same experimental techniques, the dynamics of polymers as well as their response to external forces \cite{BustamanteNature2003, MarkoSiggia1995Macromol, gittes1998dynamic, PhysRevLett.77.306, winkler2006intramolecular} or flow fields \cite{Chu1997PhysRevE, Chu1997Science, LeDucWirtz1999Nature, Winkler2006prl} may be analyzed and are now well characterized theoretically over 
a broad 
range of polymer stiffnesses. The effect of bending stiffness on the relaxational dynamics in quiescent solution \cite{leGoffFrey2002prl, Everaers2010EuropPhysJE, gittes1993jcellbiol, ISI:000220055400062, kroy1997dynamic}, and the linear response to weak external forces has been investigated in detail \cite{Morse1998MacromolViscoelast,morse1998viscoelasticity, gittes1998dynamic}. Even the response to strong fields and the ensuing nonequilibrium dynamics is fairly well understood \cite{hallat, hallatschek2007tension, *hallatschek2007tensionp2, obermayer2007stretching, obermayer2009freely, thuroff2011longitudinal}.

Here we study the dynamics of single polymers in shear flow which leads to a \emph{tumbling} motion, \emph{i.e.}\ an end-over-end turning of the polymer. It has been experimentally studied mainly for DNA~\cite{Chu1999Science, LeDucWirtz1999Nature, doyle2000dynamics, Teixeira2005Macromol,  PhysRevLett.95.018301,  Gerashchenko2006prl, Shaqfeh2005JNonNewtFluidMech}, and more recently also for F-actin~\cite{harasim2013direct}. 
There are two characteristic time scales, the shear rate $\gamma$ and the polymer's relaxation time $\tau$. Hence one expects that their ratio, known as the \emph{Weissenberg number} $W\!i = \gamma \tau$, is an important dimensionless quantity. However, there is an ambiguity in the definition of the Weissenberg number, as it remains elusive whether the relaxation time refers to 
global rotation of the polymer or internal relaxation of the segments relative to each other.  For long DNA, much longer that its persistence length $l_p$, the characteristic tumbling frequency was found to scale as a power law $W\!i^{2/3}$~\cite{Teixeira2005Macromol,  PhysRevLett.95.018301, Gerashchenko2006prl}, where $\tau$ was taken as the internal relaxation time. This is in accordance with theoretical work~\cite{ISI:000230051000012,Winkler2006prl, PhysRevE.81.041807} and numerical simulations for flexible polymers~\cite{CelaniTuritsyn2005,ISI:000232727200002, winklergompper2011, lamura2012semiflexible, munk2006dynamics, ISI:000227448200059, ISI:000088080100003}. 
It is commonly argued that the relaxation time in the Weissenberg number should be that of the slowest modes~\cite{winklergompper2011, PhysRevLett.95.018301, ISI:000165585000036}. While for a flexible polymer these clearly are internal modes, it will eventually become the global rotation with increasing polymer stiffness. In fact, recent work for 
short DNA segments and F-actin employ the rotational relaxation time~\cite{harasim2013direct, lee2011effect}. Strikingly, the tumbling of a stiff rod, which may be solved exactly, shows the same $W\!i^{2/3}$ scaling behavior as a flexible polymer~\cite{jeffery1922motion, bretherton1962, Takamura1981} despite the fact that now $\tau$ refers to the global rotation time. 
This agreement in the scaling behavior of the flexible and the stiff limit is odd as the physics of the tumbling process is qualitatively different. While flexible polymers stretch and coil up during tumbling avoiding large shear gradients~\cite{LeDucWirtz1999Nature, Gerashchenko2006prl, Chu1999Science, Teixeira2005Macromol, PhysRevLett.95.018301, aust2002rotation}, rigid rods rotate (Jeffery orbits) and are thereby exposing the full contour to shear~\cite{jeffery1922motion, harasim2013direct}. 

Cytoskeletal filaments like F-actin and microtubules are intermediate between these two extremes. Due to their finite bending stiffness they neither remain completely straight nor do they fully coil up. Indeed, recent experiments on F-actin in shear flow show that they follow a unique U-shaped path during the tumbling event, with most of the contour staying straight and the polymer exploring only a short distance, when compared to the contour length, in the shear gradient direction \cite{harasim2013direct}. 
While these experimental results have still been interpreted within the classical $W\!i^{2/3}$ scaling behavior, theoretical work clearly indicates deviations from this scaling behavior~\cite{munk2006dynamics}.

Here we perform Brownian dynamics simulations of single linear and ring polymers over a broad range of polymer stiffnesses and shear rates. We recover previous scaling results obtained in the limits of a flexible linear polymer and  a stiff rigid rod. In the regime relevant for semiflexible biopolymers we identify new scaling regimes and rationalize those in terms of two qualitatively distinct Euler buckling instabilities which lead to two types of tumbling regimes with a characteristic sequence of polymer shapes. 
We show that in addition to the Weissenberg number the stiffness of the polymer is crucial to fully characterize the tumbling behavior, and find in our simulations that for intermediate stiffness there exists a new scaling regime, where the characteristic tumbling frequency scales with a power law $\gamma^{3/4}$ distinct from the classical result. Our results quantitatively explain recent experimental results on F-actin~\cite{harasim2013direct}, both with respect to the magnitude of the tumbling frequency and the scaling with 
shear strength.  Moreover, we will also discuss the behavior of ring polymer in shear flow. These polymers are much less studied than linear polymers. Recent experimental and theoretical studies have mainly focused on equilibrium conformations \cite{alim2007shapes, Dogic2010PRL,Claessens2006}. The results for ring polymers are similar to those for linear polymers but with different power laws. In addition, to tumbling motion we also find tank-treading dynamics similar to recent studies \cite{chen2013tumbling, chen2013effects}.
For illustration we have added movies of ring and linear polymers in shear flow in the Supplemental Material \cite{supplement}. 

The paper is organized as follows: In Sec.~\ref{sec:langevin} we introduce the wormlike chain model, the Langevin equation in the free draining limit, and give a concise discussion of the numerical algorithm used for the simulations. In the following Section, we discuss the relaxation behavior of a polymer in equilibrium. We review previous results for linear polymers, and derive analytic expression for the relaxation of ring polymers. 
In Sec.~\ref{sec:shear_linear_polymer} we present a numerical study of the tumbling dynamics of a linear polymer in shear flow with special focus on the effects of polymer stiffness. Then we generalize our study to ring polymers in shear flow in Sec.~\ref{sec:ringinshear}. The appendix contains an elaborate analysis  of the numeric algorithm used to simulate ring polymers, the simulation parameters, and the analytic calculation of the relaxation dynamics of ring polymers.

\section{Langevin dynamics of a semiflexible polymer} 
\label{sec:langevin}

We describe the polymer's contour in terms of a continuous, inextensible space curve $\mathbf{r} (s,t)$, where $s$ denotes the arc length position $s \in [0,L]$. The bending energy costs for a particular polymer conformation are given by the wormlike chain Hamiltonian~\cite{ISI:A19678929500026,  ISI:A1949UX12900006} 
\begin{eqnarray}
{\cal H}[\mathbf{r}(s,t)]=\frac{\kappa}{2} \, \int_0^L \mathrm{d}s \left( \frac{\partial^2 \mathbf{r}(s,t)}{\partial s^2} \right)^2 \, .
\end{eqnarray}
Here, the bending stiffness $\kappa$ is related to the persistence length by $l_p= \kappa / k_B T$, which measures the distance over which the orientation of the tangent vectors are correlated. We are interested in the full stiffness range covering stiff polymers ($l_p \gg L$) as well as highly flexible polymers ($l_p \ll L$). 
We assume that the dynamics of the polymer in an external fluid velocity field  $\boldsymbol{u}(\boldsymbol{r})$ is  governed by a Langevin equation in the free draining limit
\begin{eqnarray}
\zeta \frac{\partial \mathbf{r}(s,t)}{\partial t} = - \frac{\delta {\cal H}[\mathbf{r}(s,t)]}{\delta \mathbf{r}(s,t)} +\zeta \boldsymbol{u} (\mathbf{r})+ \boldsymbol{\eta}(s,t) \, , 
\end{eqnarray}
where $\zeta$ denotes the friction coefficient, and $\boldsymbol{\eta}$ is a Gaussian white noise with average zero, and an amplitude determined by the Einstein relation
\begin{eqnarray}
\langle \eta_i(s,t) \, \eta_j(s',t') \rangle 
= 2   \, \zeta\, \delta_{ij} \, k_B T  \, \delta(s-s') \, \delta(t-t') \, .
\end{eqnarray}

Using the free draining limit is well justified in the stiffness range $l_p \gtrsim L$: Evaluating the Fourier transformation of the Green's function for a hydrodynamic force field (Oseen tensor)  gives only a weak (logarithmic) mode dependence of the mobility \cite{frey1991dynamics, kroy1997dynamic}. This is basically due to the mostly straight conformation of stiff filaments. 
Since the tumbling of a flexible polymer under shear is dominated by relaxation processes where the polymer is rather elongated~\cite{ISI:000232727200002}, one expects long-ranged hydrodynamic interactions to be of minor importance even in the highly flexible limit. Indeed, recent investigations show that the free draining limit yields basically the same behavior as simulations fully accounting for long-ranged hydrodynamic interactions; the tumbling frequencies are slightly overestimated~\cite{ISI:000227448200059, ISI:000088080100003}. 
These effects become even smaller upon using a dimensionless representation~\cite{Winkler2006prl}. Taken together, we conclude that it is well justified to use the free draining limit for the tumbling dynamics under shear flow over the whole stiffness regime. Note that for similar reasons we have also assumed isotropic friction in Eqs.~2 and 3.

For our numerical simulations of the Langevin dynamics we employ a \emph{bead-rod} algorithm~\cite{Morseconstrainedtheory,ISI:A1995QD78900017, ISI:A1978FP21600029} following closely the method described in Ref.~\cite{ISI:000227372200078}. For the readers convenience we give a concise summary of the basic ideas next, and refer the interested reader to Ref.~\cite{ISI:000227372200078} for an in-depth exposition of the numerical algorithm: The polymer is discretized into $N=L/b$ rods of fixed length $b$ leading to the discretization of the continuous expressions for the bending energy and equation of motion as described in Ref.~\cite{ISI:000227372200078}. 
The length constraint is implemented by Lagrangian multipliers, and additional metric pseudo-forces are introduced to make sure that the constrained dynamics yields the proper equilibrium distributions; the metric forces are implemented by an efficient algorithm introduced in Ref.~\cite{effalgometric}. 
Due to the constraints it is advantageous to interpret the stochastic differential equation using a kinetic stochastic integral \cite{ISI:000073858100006}, which implies that a specific mid-step algorithm has to be used \cite{Morseconstrainedtheory,ISI:000227372200078, ISI:000073858100006}.

In detail, this mid-step algorithm is implemented as follows: First the bond vectors for the current contour and the resulting constraints are determined. Then properly scaled noise is generated. In order to achieve a fast algorithm, we use uniformly distributed random numbers for the noise, which has been shown to amount to the same behavior as Gaussian white noise within the statistical errors \cite{ISI:A1995QD78900017}. 
Next, the noise is projected on the subspace allowed by the constraints on the bond length. Then the metric potential combined with the bending forces is calculated. Finally, noise and flow forces are added to the bending forces, and the sum of forces is projected on the allowed subspace.  With these forces a mid-step position is calculated. Now as the special feature of the required mid-step algorithm, all deterministic forces have to be evaluated a second time, using the virtual contour at the mid-step position. These forces are added to the original projected noise and projected on the 
allowed subspace to determine the actual move of the polymer in this time-step.

For the simulation of a ring polymer, one needs the additional constraint $r(t,0)=r(t,L)$. While the matrix involved in determining the Lagrangian multipliers for the projection is symmetric tridiagonal for a linear chain \cite{Morseconstrainedtheory,ISI:000227372200078}, it becomes a cyclic symmetric matrix for a ring polymer. Fortunately, cyclic matrices may be solved as efficiently as tridiagonal matrices by standard recursions and thus the calculation of the projection steps required no significant adjustment.  
The efficient calculation of metric pseudo-forces uses the same matrix as the projection and has to be extended for the ring case.  To keep the algorithm, \emph{i.e.}\ the computation time, linear in the number of beads we applied basic matrix transformations to rewrite the matrices in the usual tridiagonal form, see Appendix~\ref{app:ringmetric} for details.  On this modified matrix we used the same algorithm as for the linear chain \cite{effalgometric}.

To minimize discretization artifacts in our simulation results, we employed an iterative scheme. For each given shear flow we used two realizations with the largest and smallest values of $l_p/L$ to check the influence of the discretization, \emph{i.e.}\ the bond-length on the results. To this end we performed test simulations using a given bond length $b$, and then repeated the simulations using a halved bond length. If the final result, e.g. the power spectrum, was significantly changed, we repeated this procedure of halving the bond length. Otherwise, we took the corresponding bond length for the production runs of our simulations.  
For the simulation in shear flow the two parameters $l_p/L$ and $\gamma$ both affect the acceptable discretization. The stronger the flow, the shorter the bonds need to be, but at the same time longer bonds, \emph{i.e.}\ a lower discretization may be used for for higher values of $l_p$. Thus the precise value had to be determined for each set separately. Due to the discretization the bond length will not remain strictly constant throughout the simulation. We adjusted the numeric time-step in order to keep the maximal error in the bond length below $2\%$ during each simulation.

To complement and check our simulations with the bead-rod algorithm, we also implemented a bead-spring algorithm where the bonds are represented by springs instead of constraints. 
For the results in both types of algorithms to be comparable, the fluctuations in the bond length in the bead-spring simulations should be comparable to the error in bond length of the bead-rod simulations. This is guaranteed by using a stiff spring. The accurate numerical simulation of these strong potentials of the springs requires time steps which are much smaller than the one used in the bead-rod simulations. Therefore the bead-spring simulation needed longer computational times for a given system than the corresponding bead-rod simulation. 
Hence we restricted the bead-spring simulations to a representative sub-set of parameter sets covering the full range. In all cases, results were in excellent quantitative agreement with the corresponding bead-rod simulation.
We also checked our simulations against the known results of the conformation dynamics of linear polymers \cite{leGoffFrey2002prl, Everaers2010EuropPhysJE}. For the ring polymers we compared our results to the Monte-Carlo simulations for the equilibrium shape \cite{alim2007shapes}. In all cases we found identical results. 

For the presentation of our numerical results we define the time scale
\begin{equation}
\tau^{*} = \frac{\zeta L^3}{2k_BT}\, .
\label{def:taustar}
\end{equation}
This corresponds to the time the center of mass of a rigid rod of length $L$ takes to diffuse its own contour length; it is a convenient measure since it is independent of the persistence length $l_p$. 
For later reference, an actin filament with a contour length of $L=10 \mu \mathrm{m}$ and a diameter of $5 \mathrm {nm}$ in a solution with a viscosity of $0.1 \mathrm {Pa \ s}$ at a temperature of $20^\circ C$ has $\tau^* \approx 1.5 \times 10^4 \mathrm{s}$. Hence for such a system, \emph{e.g.}\ a flow with $\gamma \tau^* = 1.5 \times 10^5$ corresponds to a shear gradient of $\gamma \approx 10 \mathrm{s}^{-1}$. 
For the actual numerical simulation we employed time and length scales such that $k_BT=\zeta=1$ and adjusted the contour length to result in the desired value of  $\tau^{*}$.

\section{Conformational dynamics}
\label{sec:EqDynamics}

\subsection{Linear polymers}

The equilibrium relaxation dynamics of linear semiflexible polymers has been studied previously and we will only give a short summary of the relevant results needed for later comparison \cite{frey1991dynamics, kroy1997dynamic, ISI:A1996TN90500006, farge1993dynamic, granek1997semi, hallat, ISI:A1953UC11300031}.

In the parameter regime where the polymer contour length $L$ is comparable or smaller than the persistence length $l_p$, longitudinal (stored length) fluctuations are negligible and only bending (undulation) modes are important. Then the Langevin equation reduces to 
\begin{eqnarray}
\label{eq:langevinrperplinear}
\zeta \frac{\partial \mathbf{r}_\perp (s,t)}{\partial t} = - \kappa \, \frac{\partial^4 \, \delta \mathbf{r}_\perp (s,t)}{\partial s^4}+ \boldsymbol{\eta}_\perp (s,t) \, , 
\end{eqnarray}
where for the noise $\boldsymbol{\eta}_\perp (s,t)$ it holds $\langle \boldsymbol{\eta}_\perp (s,t)  \rangle =0$ and $\langle \boldsymbol{\eta}_\perp (s,t) \, \boldsymbol{\eta}_\perp (s',t') \rangle = 4   \, \zeta \, k_B T \, \delta(s-s') \, \delta(t-t') \, $. This equation can be solved by a linear mode analysis \cite{granek1997semi,hallat}. For free ends one finds for the mean square fluctuations of the end-to-end vector $\mathbf{R}$  \cite{leGoffFrey2002prl}
\begin{eqnarray}
\delta \mathbf{R}^2 := \langle (\mathbf{R}(t)-\mathbf{R}(0))^2 \rangle  = \frac{L^4}{90 l_p^2} \, F (t/ \tau_L^{\text{lin}}) \, ,
\end{eqnarray}
where 
\begin{eqnarray} 
\tau_L^{\text{lin}}=  \frac{\zeta }{\kappa} \, \left( \frac{L}{A} \right)^4 \, 
\end{eqnarray} 
with $A \approx 4.73$. While for times much smaller than the longest relaxation time $\tau_L^{\text{lin}}$, one gets a simple power law
\begin{eqnarray}
\delta \mathbf{R}^2  \cong 2.71 \, \frac{L}{l_p^2} \, \left( \frac{\kappa}{\zeta} \, t \right)^{3/4}
\end{eqnarray}
it saturates at the equilibrium value 
\begin{eqnarray}
\delta \mathbf{R}^2 = \frac{L^4}{90 l_p^2} 
\end{eqnarray}
for $t \gg \tau_L^{\text{lin}}$. Additionally, the polymer undergoes global rotation, which is purely diffusive and usually only seen after the internal relaxation saturated.

\subsection{Ring polymers}

In the following we present the first discussion of the relaxation behavior of a semiflexible ring polymer. For this we need to slightly modify the above standard approach. A convenient observable, which includes fluctuations of the size as well as the orientation of the ring, is the diameter vector of the ring defined by
\begin{eqnarray} 
 \mathbf{D}(t) := \mathbf{r}(0,t) - \mathbf{r}(\frac{L}{2},t) \, 
\end{eqnarray} 
where the choice of $s=0$ is arbitrary. To calculate the mean-square displacement of the ring diameter, $\delta \mathbf{D}^2 (t) = \langle (\mathbf{D}(t)-\mathbf{D}(0))^2 \rangle$, we assume that the semiflexible polymer's configuration is effectively constrained to a plane \cite{alim2007shapes}. Then, the polymer dynamics may be decomposed into global rotation and internal relaxation of this planar polymer configuration. 
As any change of the plane requires the whole ring to move, this will happen on a much larger time scale than the relaxation within the plane, and we may separate these two processes, such that $\delta \mathbf{D}^2=\delta \mathbf{D}^2_{\mathrm{rot}}+\delta \mathbf{D}^2_{\mathrm{shape}}$. The rotation of the configuration plane is equivalent to a random walk on a unit sphere \cite{ISI:A19678929500026}, and hence
\begin{equation} 
\delta \mathbf{D}^2_{\mathrm{rot}} = 8R^2 \left[1- \exp \left(- t / \tau_L \right) \right],
\label{eq:msd-rotation}
\end{equation}
where $R=L/2\pi$, and
\begin{eqnarray} 
\tau_L =  \frac{\zeta L^3}{16 \pi^2 k_BT} 
\end{eqnarray} 
denotes the longest rotational relaxation time. The internal relaxation within the plane, which we assume to be two-dimensional, may be determined using the weakly bending limit and is calculated similarly to the familiar rod-shaped case \cite{leGoffFrey2002prl, kroy1997dynamic}. See Appendix~\ref{app:ringanalytisch} for details on the calculations. We find
\begin{equation} 
 \delta \mathbf{D}^2_{\mathrm{shape}} = \frac{8R^3}{\pi l_p} \sum_{n\ \mathrm{odd}} \frac{1}{n^4+1} \left[ 1- \mathrm{e}^{- (n^4+1) t / \tilde \tau_L } \right]
\label{eq:msd-shape}
\end{equation} 
with the longest internal relaxation time given by 
\begin{eqnarray} 
\tilde \tau_L = \frac{\zeta}{\kappa} R^4 \, .
\end{eqnarray} 
Thus, already for moderate filament stiffness, $l_p \ge L/ \pi^2$, the longest relaxation time is determined by rotation. 

\begin{figure}[h]
      \begin{center}
		\includegraphics[width=0.9\columnwidth]{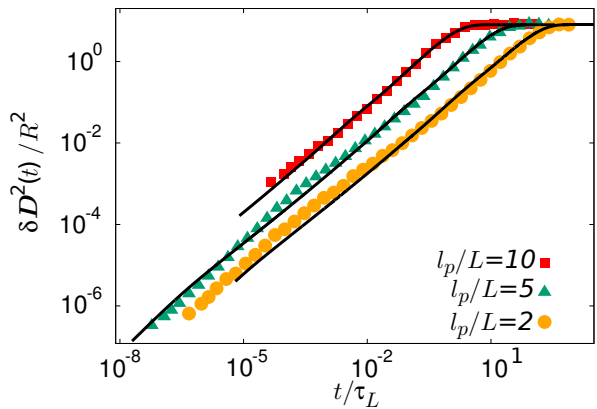}
     \end{center}
      \caption{(Color online) Time evolution of the mean-square displacement of the diameter of cyclic polymers, $\delta \mathbf{D}^2$, for different polymer stiffness $l_p/L$ indicated in the graph. For better visibility, the curves with $l_p/L=5$ and $l_p/L=2$ are shifted by a factor of $10$ and $100$, respectively.
      Symbols give the numerical data, standard error is below symbol size. Solid lines give the corresponding analytic prediction as obtained from Eq.~\ref{eq:msd-rotation} and Eq.~\ref{eq:msd-shape}. \label{fig:RelaxationNumAnaly}}
\end{figure}

To test our analytic calculations we determined the mean-square displacement of the diameter $\delta \mathbf{D}^2$ using our Brownian dynamic simulations, and find good agreement, as can be inferred from Fig.~\ref{fig:RelaxationNumAnaly}. The small deviations near the cross-over from internal to rotational relaxation we attribute to the assumption of decoupled global and internal relaxation modes.

\begin{figure}[htb]
  \begin{center}
		\includegraphics[width=8.0cm,height=5.5cm]{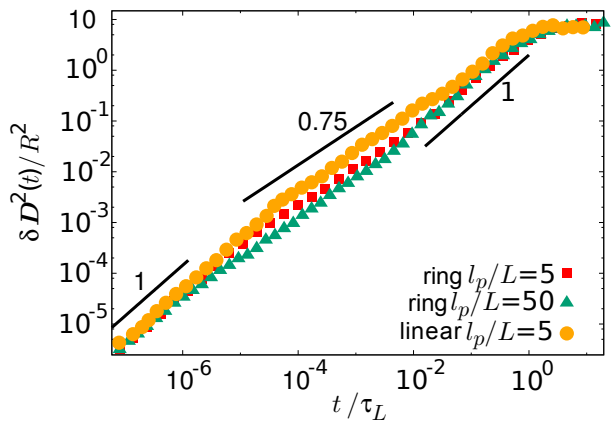}
     \end{center}
\caption{(Color online) Mean-square displacement of the ring diameter, $\delta \mathbf{D}^2$,  as a function of time for two ring polymers with stiffness $l_p/L=5$ and $l_p/L=50$, respectively.  As predicted from the calculations, the stiffness does not affect the rotational relaxation time nor the saturation value. 
For comparison the mean-square displacement of the end-to-end vector of a linear polymer is shown, and we adjusted the contour length of the linear chain $L_l$ to be identical to the diameter of the ring polymer, \emph{i.e.}\ $L_l=L/\pi$. For this linear polymer  $l_p/L_l=5$.
Statistical errors of the Brownian dynamics simulations are of the order of the symbol size. 
\label{Relaxationregime}}
\end{figure}

In Fig.~\ref{Relaxationregime} 
we compare the conformational dynamics of ring polymers and linear polymers. In order to get identical saturation values of the mean-square displacement of the end-to-end vector and the diameter, respectively, at asymptotically large times we take the length of the linear polymer to be equal to the ring diameter. With this choice the longest, rotational relaxation time of the linear polymer is slightly shorter than $\tau_L$ of the ring polymer.
The initial relaxation is identical and linear in $t$ since it corresponds to free diffusion of individual beads. At intermediate times, both linear and ring polymers exhibit a $t^{3/4}$-scaling, consistent with experimental data on linear F-actin filaments \cite{leGoffFrey2002prl}. However, the relaxation time and the amplitude of the internal modes for ring polymers are smaller by a factor of approximately $2$. 
This is consistent with the reduced fluctuations of F-actin rings reported recently~\cite{Dogic2010PRL}. Intermediate between the $t^{3/4}$-scaling and the final plateau one observes a  linear diffusive regime due to rotational motion, $\delta \mathbf{D}^2_{\mathrm{rot}}$. As a consequence of the smaller internal relaxation times, this regime begins at earlier times, and is hence more pronounced for ring polymers as compared to linear polymers. 
The crossover times are proportional to $L^4/l_p$ and $b^4/l_p$, respectively. Taken together, linear and ring polymers show similar crossover behavior in the conformational dynamics. The ring topology only affects prefactors in the amplitude and the crossover time scales.

\section{Linear polymers in shear flow}
\label{sec:shear_linear_polymer}

\begin{figure}[htb]
\begin{center}
  \includegraphics[width=6cm,height=6cm]{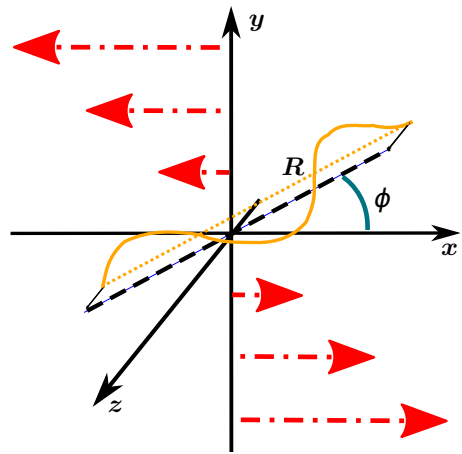}
\end{center}
\caption{(Color online) Schematic representation of the flow geometry for a linear
polymer in linear shear flow  $\mathbf{v}= \gamma \, y \,
\hat{\mathbf{e}}_x $ [dashed-dotted red (dark gray) arrows]. The inclination of the polymer's
end-to-end distance $\mathbf{R}$ [dotted yellow (light gray) line] with respect to the
neutral plane ($xz$-plane) is measured in terms of the angle $\phi$
between the projection of $\mathbf{R}$ onto the shear plane ($xy$-plane)
[dashed line] and the $x$-axis.
\label{def_phi}}
\end{figure}

\begin{figure}[htb]
	\begin{center}
               \includegraphics[width=0.9\columnwidth]{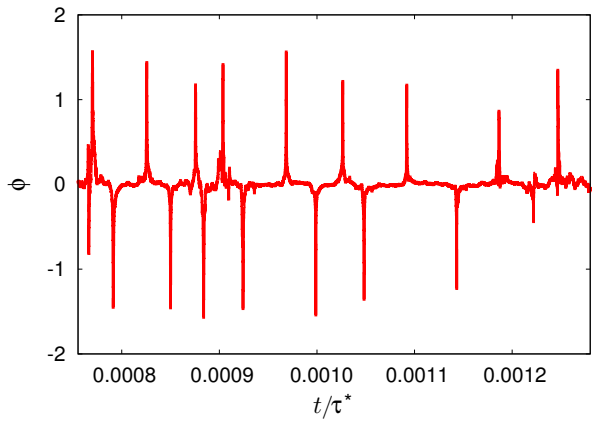}
	\end{center}
     \caption{(Color online) Typical time evolution of the angle $\phi$ for a linear polymer with $l_p/L=2$  in shear flow with $\gamma \tau^{*}=1.35 \times 10^4$. \label{timeev_linear}}
\end{figure}

\begin{figure}[htb]
  \begin{center}
    \includegraphics[width=\columnwidth]{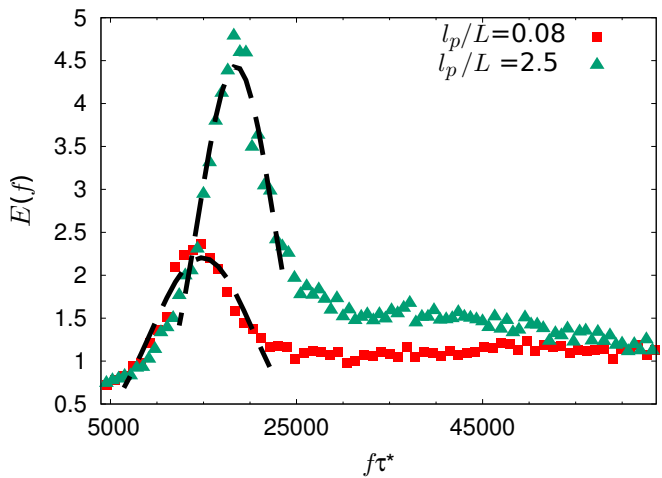}
  \end{center}
\caption{(Color online) Typical power spectra $E(f)$ for the angle $\phi$ at a shear
rate $\gamma \tau^* = 5 \times 10^6$ and for two different values of the
polymer stiffness as indicated in the graph. The dashed lines show
Gaussian fits to the central region of the peak at the tumbling
frequency $f_c \tau^*$.
 \label{fig:linpsd}}
\end{figure}

In this section we present the results of our Brownian dynamics simulations of linear polymers in shear flow 
\begin{equation}
\mathbf{v}= \gamma \, y \, \hat{\mathbf{e}}_x \, ,
\end{equation} 
where  $\gamma$ is the shear rate. The flow geometry is illustrated in Fig.~\ref{def_phi}. In this geometry, a polymer fully embedded in the $x z$-plane is not subject to any forces from the shear flow; hence we call this the \emph{neutral plane}. 
However, any thermal fluctuations will inevitably lead to polymer conformations which are inclined with respect to the neutral plane. We measure this inclination by the angle $\phi$ between the projection of the end-to-end vector $\mathbf{R}$ onto the $xy$-plane (shear plane) and the $x$-axis~\cite{ISI:000230051000012, ISI:000232727200002, Gerashchenko2006prl, Winkler2006prl}. Shear forces then lead to a tumbling motion of the polymer. A typical time trace of $\phi$ consists of extended time periods where the polymer shows small fluctuations close to the neutral plane which are interrupted by fast tumbling events, cf. Fig.~\ref{timeev_linear}. 
From these time traces we calculated the power spectral density by using the Wiener-Khinchin theorem \cite{gardiner1985handbook}. To this end long time traces extending over $2000$ to $100 \, 000$ tumbling events were recorded, and used to first determine the autocorrelation function $\langle \phi(t) \phi(t-t_k) \rangle$ at $n=400$ equidistant points in time $t_k$, where the brackets denote a moving time average, averaging over $t$ for fixed $t_k$~\footnote{The most prominent feature of the power spectral density is a peak at a frequency corresponding to the tumbling rate. 
We have optimized the location and distance between the time points $t_k$ to resolve this peak well. To get an initial value for $t_k$, we tested various values for $t_k$ for two simulations with $l_p/L= 0.4$ and $l_p/L=10$ and $\gamma \tau^{*} \approx 4.1 \times 10^{4}$ to determine the shape of the power spectrum. A good representation, \emph{i.e.}\ including the decline at both sides of the peak over at least a factor of $5$, was found for $t_k/\tau^{*}=9.7 \times 10^{-5}$. 
For all further simulations, we scaled $t_k$ with $(\gamma \tau^{*})^{2/3}$ relative to this starting value. In each particular simulation slight further adjustments of the time points $t_k$ were used whenever necessary to obtain a clear decline at both sides of the peak in the power spectral density.}. Next, upon taking the discrete Fourier transformation of the autocorrelation function the power spectral density was obtained
\begin{eqnarray}
  E(f)= \left| \sum_{k=0}^{n-1} \langle \phi(t) \phi(t-t_k) \rangle \exp(-2\pi i f t_k) \right| \, .
\end{eqnarray}
As illustrated in Fig.~\ref{fig:linpsd}, these power spectra show a pronounced peak at some frequency $f_c$ which can be taken as a good proxy for the characteristic tumbling rate\footnote{In experimental studies frequently an alternative approach for the definition of the tumbling frequency is used. The number of turning events is recorded either directly or determined by using the unfolded angle $\phi$, \emph{i.e.}\ recording the total covered angular distance, which then is divided by $\pi$. The frequency may then by calculate by dividing the number of events by the total time under consideration. Throughout the full range of $\gamma \tau^*$ under consideration here both definitions agree well within the numerical errors, see \emph{e.g.}\ Fig.~\ref{fig:linexperiment}.}. We used Gaussian fits to determine the peak positions $f_c$ of the power spectra as illustrated in Fig.~\ref{fig:linpsd}. 

\begin{figure}[htb]
     \begin{center}
		\includegraphics[width=\columnwidth]{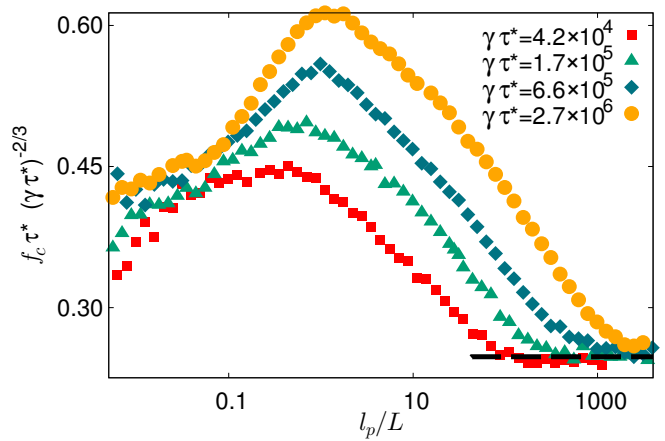}
     \end{center}
  \caption{(Color online) Scaling plot showing ${f_c \tau^{*}}/(\gamma \tau^{*})^{2/3}$ as a function of $l_p/L$ for a set of values for the shear rate $\gamma \tau^{*}$ as indicated in the graph. The dashed line in the stiff limit is the tumbling frequency of a rigid rod derived from Jeffery's equation~\cite{jeffery1922motion, bretherton1962}. This Jeffery plateau is reached the later the stronger the shear flow. }
\label{linmtfvonlp-classical}
\end{figure}

Figure~\ref{linmtfvonlp-classical} shows the dimensionless characteristic tumbling frequency $f_c \tau^{*}$ as a function of the relative stiffness $l_p/L$ for a set of dimensionless shear rates $\gamma \tau^{*}$ indicated in the graph. Here we have rescaled $f_c \tau^{*}$ with $(\gamma \tau^{*})^{-2/3}$ as suggested by previous theoretical results~\cite{PhysRevLett.95.018301, Gerashchenko2006prl, Teixeira2005Macromol, CelaniTuritsyn2005,Winkler2006prl, winklergompper2011, PhysRevE.81.041807, lamura2012semiflexible}.  
From Fig.~\ref{linmtfvonlp-classical} we infer that both for highly flexible ($l_p/L \lesssim 0.1$) and for almost stiff polymers ($l_p/L \gg1$), $f_c $ obeys a $(\gamma \tau^{*})^{2/3}$-scaling law quite well as indicated by the data collapse. The slight deviation from a perfect collapse in the flexible limit is attributed to an increased numerical errors in this regime resulting from an increased dependence on the discretization and the small values of the end-to-end vector. Note also that the data collapse is better for large values of $\gamma \tau^*$. 
Moreover, in both of these asymptotic regimes $f_c \tau^{*}$ is largely independent of the polymer's stiffness $l_p/L$ in accordance with previous theoretical work~\cite{Winkler2006prl, harasim2013direct}. Deviations are found only at low shear rate and hence small $W\!i=\gamma \tau^*$ where $f_c$ shows a pronounced downturn for small $l_p/L$; this does not contradict existing theories since they are strictly valid only for high $W\!i$. For stiff 
filaments our simulations are in full accord with the theoretical value from Jeffery's theory~\cite{jeffery1922motion, bretherton1962, harasim2013direct} [dashed line in Fig.~\ref{linmtfvonlp-classical}]
\begin{equation}
f_c \tau^* = \frac{3 \times 6^{1/3}}{22}  \left( \gamma \tau^* \right)^{2/3} \, .
\end{equation}
There is, however, a broad intermediate stiffness regime, covering several orders of magnitude, where the characteristic tumbling frequency $f_c$ is neither independent of the persistence length nor does it follow a $(\gamma \tau^{*})^{2/3}$-scaling law. It rather exhibits a non-symmetrical peak whose position shifts to larger values of $l_p/L$ with increasing shear rate $\gamma \tau^{*}$, and with it the asymptotic approach to the Jeffery plateau is shifted towards larger polymer stiffnesses the stronger the shear flow.

\begin{figure}[htb]
     \begin{center}
		\includegraphics[width=\columnwidth]{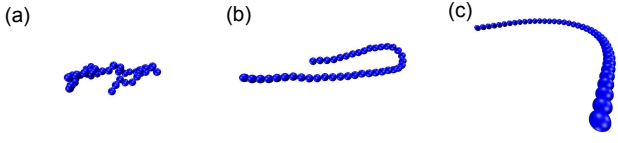}
     \end{center}
  \caption{(Color online) Typical examples for the different types of tumbling, depending on which type of Euler buckling occurs. For all examples shown here $\gamma \tau^* = 8 \times 10^4$. (a) For a flexible polymer with $l_p/L=0.025$ the polymer coils up during the tumbling. (b) A polymer with an intermediate stiffness of $l_p/L=2$ exhibit localized bends (hairpin configurations) during tumbling. (c) Finally, for stiff polymers (here  $l_p/L=80$) the polymer configurations are weakly curved over their full length.}
\label{fig:typesoftumbling}
\end{figure}

Both of these features of the tumbling frequency can be attributed to the interplay between shear flow and bending modes. In fact they correspond to two qualitatively different tumbling regimes due to distinct types of Euler buckling instabilities in shear flow, which we term \emph{local} and \emph{global} Euler buckling, as illustrated in Fig.~\ref{fig:typesoftumbling}. 
Consider a linear polymer in shear flow with a fixed value of $\gamma \tau^*$, \emph{i.e.}\ fixed shear rate and polymer length. An almost stiff polymer will rotate like a rigid rod and perform Jeffery orbits. However, upon decreasing the persistence length, at some point the shear flow will be strong enough to overcome the Euler buckling force of the polymer, $F_e  \sim l_p / L^2$, and the polymer as a whole will bend during a tumbling event; this is indeed observed in our simulations, see Fig.~\ref{fig:typesoftumbling}c.
\begin{figure}[htb]
     \begin{center}
		\includegraphics[width=\columnwidth]{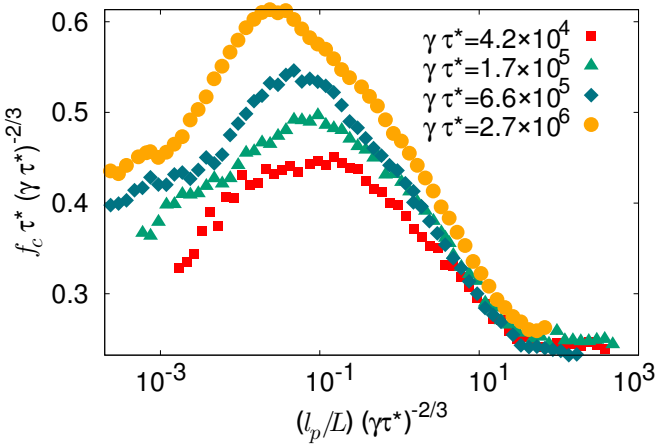}
     \end{center}
  \caption{(Color online) Modified scaling plot for the tumbling frequency to achieve data collapse for the stiff to rigid regime $0.01 \, (\gamma \tau^{*})^{2/3} \lesssim l_p/ L \lesssim 1 \, (\gamma \tau^{*})^{2/3}$. The rescaled tumbling frequency $ f_c \, \tau^{*} ( \gamma \, \tau^{*})^{-2/3}$ is shown as a function of $(l_p/L) (\gamma \tau^*)^{-2/3}$ for a series of shear rates $\gamma \tau^{*}$ indicated in the graph.}
\label{fig:crosstorigid}
\end{figure}
The threshold value of the persistence length, where the polymer starts to buckle, is determined by balancing the mechanical Euler buckling force and the shear force. The shear flow exerts a force $\gamma y  \times \zeta L$, where we may take for $y$ the typical transverse displacement caused by thermal forces: $y \sim r_\perp \sim \sqrt{L^3/l_p}$~\cite{leGoffFrey2002prl}. 
This implies for the threshold value of the persistence length: $l_{p}/L \sim (\gamma \tau^*)^{2/3}$. Rescaling the data for the tumbling frequencies according to this scaling argument results in data collapse for the onset of the stiff regime (Jeffery limit), cf. Fig.~\ref{fig:crosstorigid}. When  the polymer becomes even more flexible, there is a second shear-strength-dependent threshold value for $l_p/L$ where local buckling on length scales comparable to thermal bending modes becomes possible: $L^2_{\text{bend}} \sim r_\perp^2 \sim L^3/l_p$. Again, balancing mechanical and shear forces, $F_e \sim l_p / L_{\mathrm{bend}}^2 \sim \gamma L_{
\mathrm{bend}} \times \zeta L_f$, with $L_f= L_{\mathrm{bend}}$, yields a threshold value $l_p/L \propto (\gamma_c \tau^*)^{1/3} $. 
As can be inferred from Fig.~\ref{linmtfvonlp}, rescaling data according to this scaling behavior gives excellent data collapse within the range of statistical fluctuations of the data.
In order to achieve good data collapse for the tumbling frequency in this intermediate stiffness regime, $0.001 \, (\gamma \tau^{*})^{1/3} \lesssim l_p/ L \lesssim 1 \, (\gamma \tau^{*})^{1/3}$, requires also to rescale  the tumbling frequency. We find the best data collapse for
\begin{equation}
f_c \, \tau^{*} \sim ( \gamma \, \tau^{*})^{3/4} \, .
\end{equation}
\begin{figure}[hbt]
	\begin{center}
		\includegraphics[width=8.0cm,height=5.5cm]{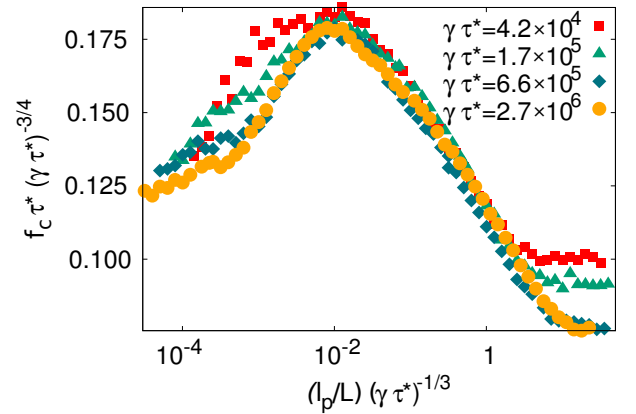}
	\end{center}
\caption{(Color online) Modified scaling plot for the tumbling frequency to achieve data collapse for the intermediate stiffness regime $0.001 \, (\gamma \tau^{*})^{1/3} \lesssim l_p/ L \lesssim 1 \, (\gamma \tau^{*})^{1/3}$. The rescaled tumbling frequency $ f_c \, \tau^{*} ( \gamma \, \tau^{*})^{-3/4}$ is shown as a function of $(l_p/L) (\gamma \tau^*)^{-1/3}$ for a series of shear rates $\gamma \tau^{*}$ indicated in the graph.
\label{linmtfvonlp}}
\end{figure}

The polymer conformation resulting from such a local Euler buckling event are U-shaped as illustrated in Fig.~\ref{fig:typesoftumbling}b; see also the videos in the Supplementary Material~\cite{supplement}. In accordance with recent experimental results~\cite{harasim2013direct} the polymer shows a specific sequence of conformations:  starting from a fully stretched state it first acquires a configuration similar to the letter J. 
The ends of the polymer then travel around a stadium track assuming the typical U-shape, and further on return to a mirrored J before it becomes fully stretched again.  As an additional theoretical insight, it was shown in~\cite{harasim2013direct} that the actual bending radius of the U-turn can be calculated by balancing shear and bending forces. This further confirms our above scaling argument for the onset of the local buckling instability which determines the value of the tumbling frequency.

\begin{figure}[tb]
	\begin{center}
		\includegraphics[width=\columnwidth]{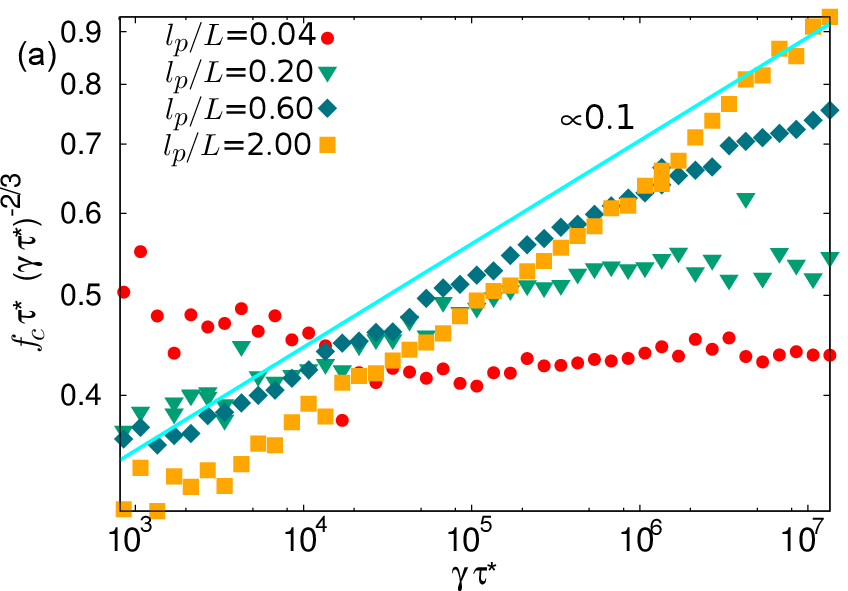}  
	\end{center}
	\begin{center}
		\includegraphics[width=\columnwidth]{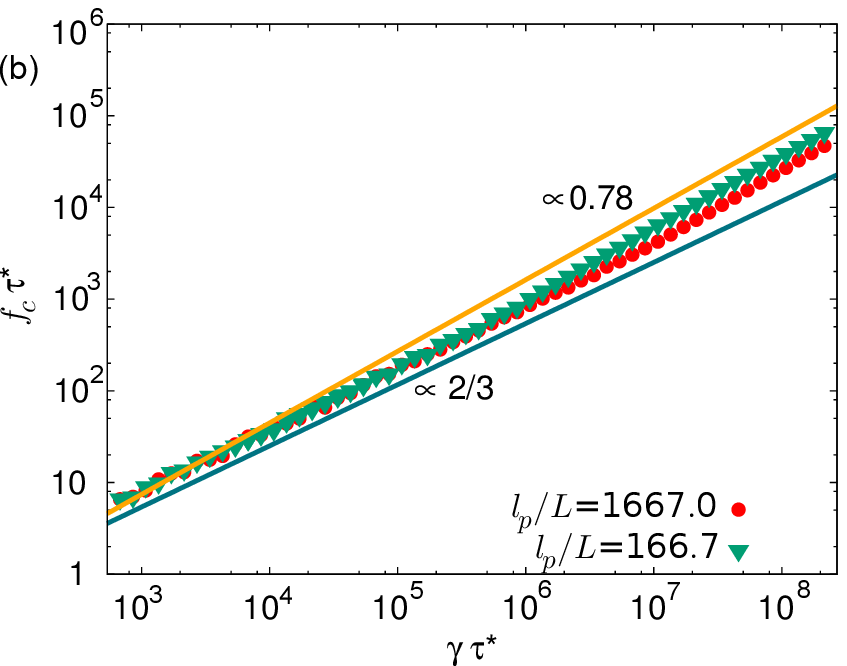}
	\end{center}
\caption{(Color online) For different values of $l_p/L$ the dependence of the
characteristic tumbling frequency on the shear rate $\gamma \tau^*$ is
determined. In (a) we show $f_c \tau^* (\gamma \tau^*)^{-2/3}$ for linear
polymer of weak to intermediate stiffness.  With increasing stiffness
an characteristic deviation of about $0.1$ occurs for small, but
increasing values of $\gamma \tau^*$.
 For $l_p/L=2$ the deviation persists throughout the whole range of
$\gamma \tau^*$ used here, which should also cover the experimentally
accessible regime. In (b) we show directly the data for $f_c \tau^*$ for
stiff polymers. The scaling exponent changes in the other direction than
for the flexible to intermediate filament, \emph{i.e.}\ the new scaling
exponent occurs at high values of $\gamma \tau^*$, which increase with
the stiffness of the polymer. \label{linmtfvongamma}}
\end{figure}

\begin{figure}[htb]
	\begin{center}
		\includegraphics[width=\columnwidth]{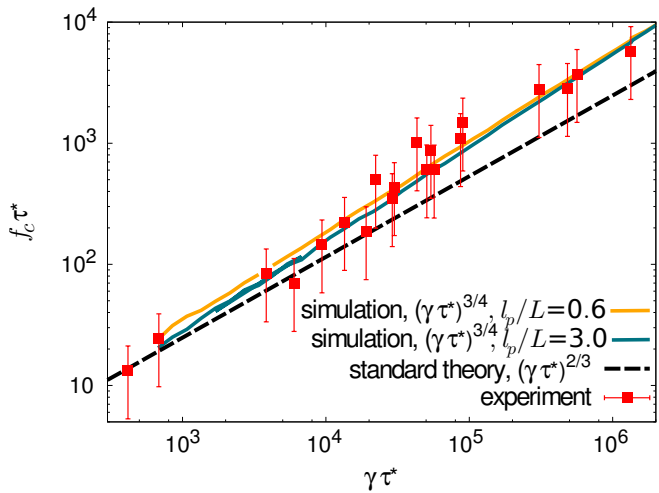}
	\end{center}
  \caption{(Color online) Comparison of our simulational results, exhibiting an increase of the frequency with $(\gamma \tau^*)^{3/4}$, to experimental data [squares in the plot] for F-actin\cite{harasim2013direct}. The experimental data agree well with the Brownian dynamics data and follow a $(\gamma \tau^*)^{3/4}$ scaling law. The data as well as the simulations are not consistent with the classical $2/3$ scaling law [solid line].  \label{fig:linexperiment}}
\end{figure}

As the exponent $3/4$ deviates significantly from the exponent $2/3$ found for both flexible polymers and rigid rods, we decided to study the shear rate dependence explicitly in order to scrutinize this startling result,  cf. Fig.~\ref{linmtfvongamma}.  
For the most flexible case under consideration we recover the predictions of the existing theories~\cite{PhysRevLett.95.018301, ISI:000227448200059, Teixeira2005Macromol, CelaniTuritsyn2005,Winkler2006prl, winklergompper2011, lamura2012semiflexible} throughout almost the full range of shear strength studied here. However, upon increasing the polymer's stiffness a different scaling regime with a larger exponent emerges over a continuously growing range of $\gamma \tau^*$. Eventually, at $l_p \approx L$, this stronger scaling with $\gamma \tau^*$ dominates over the full range of shear rates, cf. Fig.~\ref{linmtfvongamma}a. 
The exponent is consistent with the above intermediate scaling regime: $f_c \tau^* \propto (\gamma \tau^*)^{3/4}$. Consequently, as experimentally accessible shear rates are limited, typically in the range of $\gamma \tau^* \lesssim 10^7$ \cite{harasim2013direct, Gerashchenko2006prl}, we expect semiflexible polymers like F-actin in a shear flow to exhibit a tumbling frequency proportional to $W\!i^{3/4}$ throughout these experimentally accessible shear rates. This actually explains the 
systematic deviation of the experimentally measured tumbling frequencies at high $W\!i$ for F-actin observed in Ref.~\cite{harasim2013direct}, as well as the numerical findings in Ref.~\cite{munk2006dynamics}. Increasing the persistence length even further to values $l_p \gg L$, we observe an inverse crossover back to $f_c \propto W\!i^{2/3}$ (Jeffery regime), cf. Fig.~\ref{linmtfvongamma}b.    

In Fig.~\ref{fig:linexperiment} we compare our simulation results with measurements of the tumbling frequency of actin filaments as a function of the shear rate $\gamma \tau^*$ and in a length range of $L=3 - 40\ \mu$m~\cite{harasim2013direct}. 
The data clearly deviate from the classical scaling law,  $(\gamma \tau^*)^{2/3}$, as obtained for both flexible polymers~\cite{PhysRevLett.95.018301, CelaniTuritsyn2005,Winkler2006prl, PhysRevE.81.041807} and rigid rods~\cite{harasim2013direct}. The dashed line shown in Fig.~\ref{fig:linexperiment} is the quantitative result as derived in~\cite{harasim2013direct}. 
In contrast, our findings based on numerical simulation agree - without any adjustable parameter - extremely well with the experiment data, not only with respect to the new power law,  $(\gamma \tau^*)^{3/4}$, but also with respect to the amplitude of the numerical data. Note also that the two solid lines were obtained from simulations with $l_p/L=0.6$ and $l_p/L=3$ showing that the dependence on the polymer stiffness is weak within the experimental range. We highly welcome experiments on other experimental model systems of semiflexible polymers such as microtubules~\cite{gittes1993jcellbiol, pampaloni2006thermal} or nanotubes~\cite{duggal2006dynamics, ericson2004macroscopic} to further test our theoretical predictions.

\section{Ring polymers in shear flow}
\label{sec:ringinshear}

\begin{figure}[htb]
\begin{center}
   \includegraphics[width=6cm,height=6cm]{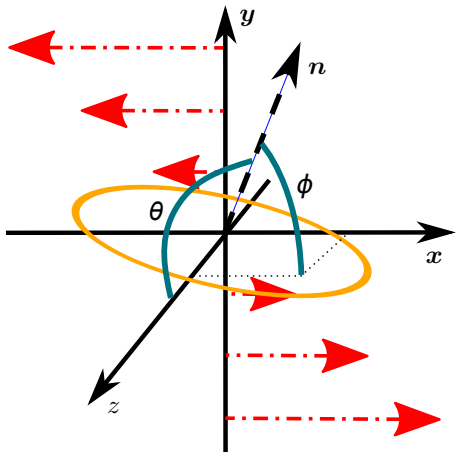}
\end{center}
\caption{(Color online) Schematic representation of the flow geometry for a ring polymer [yellow (light gray)] in linear shear flow  $\mathbf{v}= \gamma \, y \, \hat{\mathbf{e}}_x $ [dashed-dotted red (dark gray) arrows]. The ring plane is characterized by its normal [dashed line]. To fully characterize the ring's orientation relative to the shear we use the angle $\phi$ between the normal and the $xz$-plane, and the angle $\theta$ between the normal and the $z$-axis.
\label{sketch_ring_polymer}}
\end{figure}

In this section we discuss the dynamics of ring polymers in shear flow. To this end we employ the same kind of simulations as described earlier for the relaxation dynamics of ring polymers, and the implementation of shear flow as for linear polymers. To monitor the dynamics of the ring we define the normal to the ``ring plane'' as
\begin{equation}
\mathbf{n}=\mathbf{D}(\alpha) \times \mathbf{D}(\alpha+L/4) \, ,
\end{equation}
where the arc length position $\alpha$ being arbitrary in principle, was chosen as $\alpha=0$. The orientation of the ring is specified by the angle $\phi$ between $\mathbf{n}$  and the $xz$-plane, similar as for linear polymers, and in addition by the inclination $\theta$ of $\mathbf{n}$ with respect to the $z$-axis, see Fig.~\ref{sketch_ring_polymer}. 

\begin{figure}[htb]
	\begin{center}
		\includegraphics[width=\columnwidth]{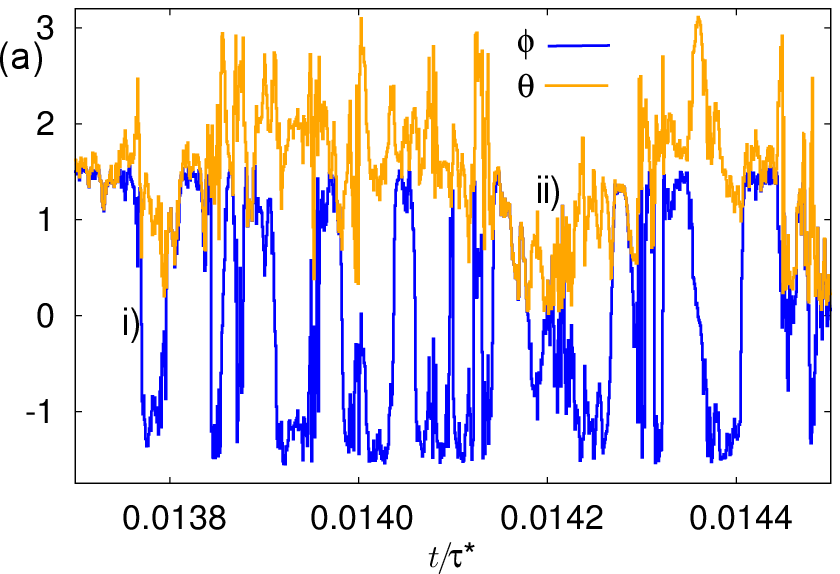}
	\end{center}
	\begin{center}
               \includegraphics[width=\columnwidth]{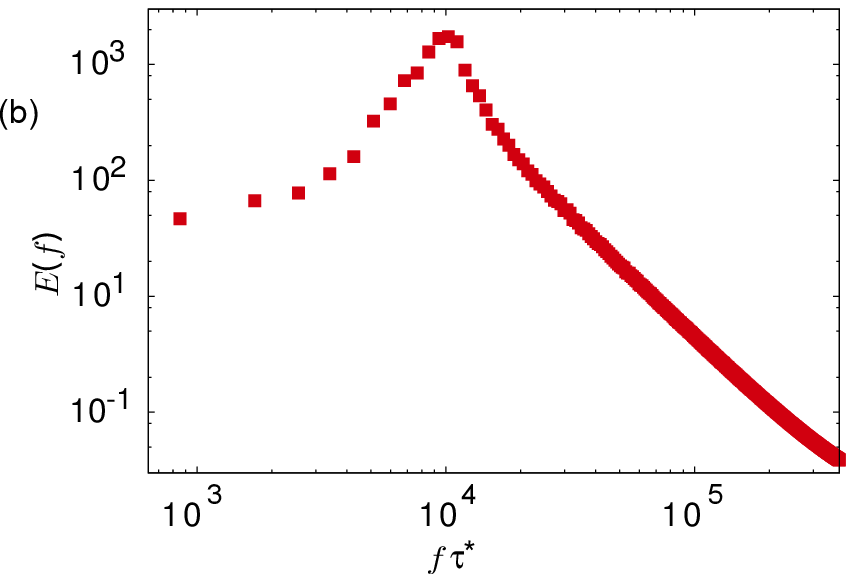}
	\end{center}
\caption{(Color online) (a) Typical time evolution of the angle $\phi$ and $\theta$ for a ring with $l_p/L=2$ under a shear flow with $\gamma \tau^*=62500$. (b) Power spectrum of the same ring polymer as in a) for the time trace of $\phi$. 
\label{timeev}}
\end{figure}

As reported in \cite{chen2013tumbling} two qualitatively different types of tumbling events may be distinguished depending on the inclination $\theta$. Imagine a ring embedded in the  $xz$-plane subject to a shear flow as indicated in Fig.~\ref{sketch_ring_polymer}. In the first type of tumbling event, the normal of the ring rotates within the $xy$-plane, such that $\theta=\pi/2$ throughout the whole tumbling event. 
A typical time trace of such a  \emph{`rapid turnover'} event is shown as $i)$ in Fig. \ref{timeev}: the turnover corresponds to a sharp change in the angle $\phi$ by about $\pi$. In the second type of tumbling event, which we term \emph{`tank-treading'}, the normal to the ring aligns with the $z$-axis ($\theta=0$ and $\phi= \pm \pi/2$).  In this configuration the ring plane coincides with the shear plane, and the shear gradient along the contour causes the ring to perform a tank-treading motion. Since the force due to the shear flow lacks a component to change the orientation of the ring plane 
this state is metastable, and 
sufficiently  strong fluctuations of the contour are needed to complete the event and return to the neutral plane.  A typical time trace of this event is shown as $ii)$ in Fig.~\ref{timeev}: the angle $\phi$ only performs fluctuations while the absolute value of $\theta$ decreases to $0$ for a time much longer than the typical duration of a rapid turnover event. These two kinds of tumbling events are also illustrated in the movies found in the Supplementary Material~\cite{supplement}.
Of course, these two are two rather idealized scenarios, and mixed tumbling events are actually rather common since the initial orientation of the ring before tumbling is broadly distributed. Hence the characteristics of tumbling events mix, e.\,g.\ tank-treading may also be observed during quite short turnover events. 

\begin{figure}[tb]
		\includegraphics[width=8cm,height=5.5cm]{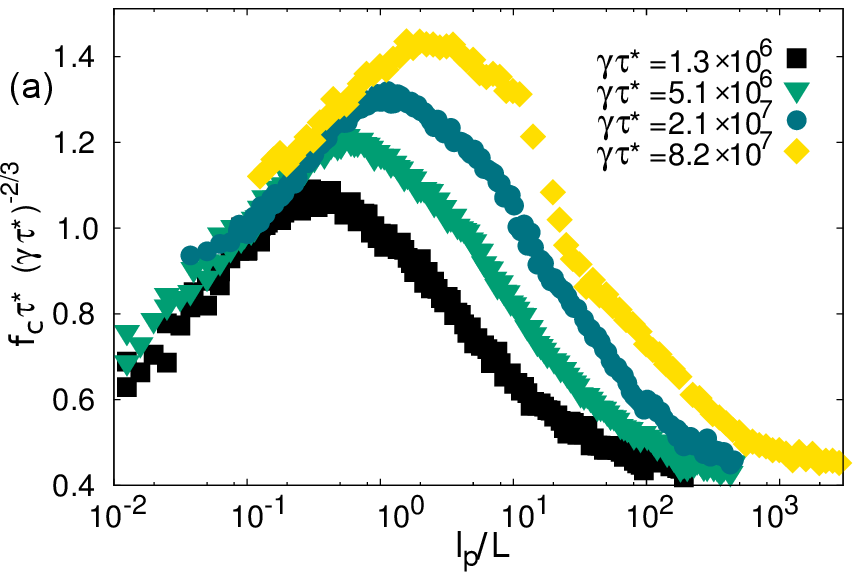}
               \includegraphics[width=8cm,height=5.5cm]{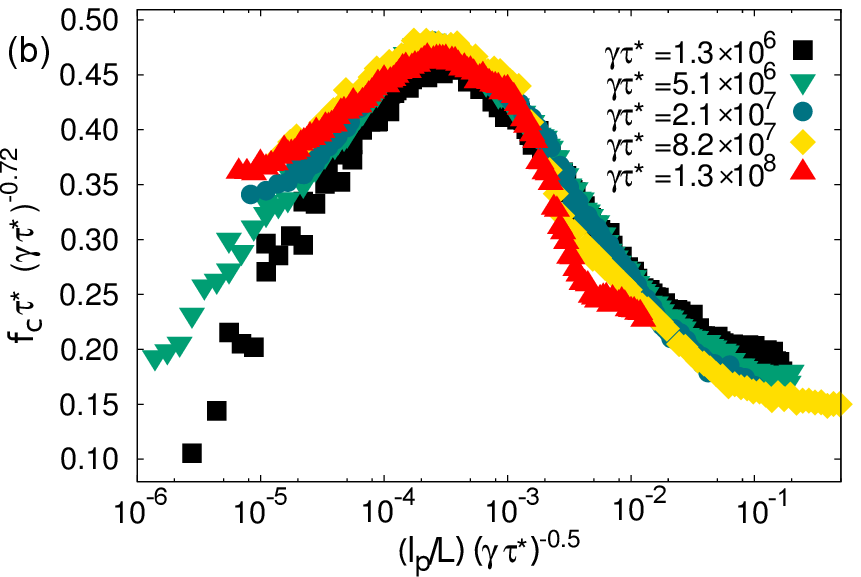}
  \caption{(Color online) Rescaled tumbling frequency $f_c$ as a function of $l_p/L$ for a ring polymer. As shown in a), rescaling only the dimensionless frequency with $(\gamma \tau^*)^{2/3}$ again results in overlapping curves in the flexible and stiff limit. In b) we show that using an exponent of $0.72$ when rescaling the frequencies and shifting the rigidity by a factor $(\gamma \tau^*)^{0.5}$ gives good agreement from the flexible regime to the peak. For higher values of $(\gamma \tau^*)$ there seem to exist small deviations in the fall to the stiff regime.
\label{ringmtfvonlp}}
\end{figure}

As for the linear polymers we analyze how the tumbling frequency $f_c$ depends on the relative persistence length $l_p/L$ for various values of the dimensionless shear rates $\gamma \tau^*$. The resulting curves are qualitatively similar to those obtained of linear polymers, see Fig.~\ref{ringmtfvonlp}. 
In the flexible limit, in the range studied here, the frequency grows linearly with $\ln(l_p/L)$. At some shear-rate-dependent stiffness the frequency starts to decline resulting in a peak structure as for linear polymers. In the stiff limit, which again starts for larger relative stiffness $l_p/L$ the higher the shear rates, there is hardly any dependence on $l_p/L$. 

Keeping the discussion analogous to the linear case, upon rescaling the tumbling frequency with $(\gamma \tau^*)^{-2/3}$ we find data collapse only in the flexible and the stiff limit, in accordance with the results in \cite{chen2013tumbling}. In contrast to the linear case, the tumbling frequency does not become constant even down to very low values of $l_p/L$. 
We are not completely sure on how to interpret this numerical observation. Most likely it indicates that for ring polymers the crossover to the fully flexible regime occurs at much lower stiffness than for linear polymers\cite{alim2007shapes, alim2007fluctuating, drube2010excluded}, as suggested by the different relaxations times as determined in eq.~14. 
In the intermediate semiflexible regime there is again no data collapse if $f_c$ is rescaled by $(\gamma \tau^*)^{-2/3}$. As for linear polymers, the curves differ primarily by a shift of the peak position to higher values of $l_p/L$ for stronger shear, and an increase of $f_c$ with increasing $\gamma \tau^*$. However, the precise numerical values of the scaling exponents are different. We find the best data collapse upon rescaling the position of the peaks by $(\gamma \tau^*)^{-0.5}$, and the frequency $f_c$ by  $(\gamma \tau^*)^{-0.72}$. 

\section{Summary and Conclusions}

We have presented a comprehensive computational analysis of the tumbling dynamics of single linear and ring polymers in linear shear flow of strength $\gamma$. To this end we have implemented a bead-rod algorithm following earlier work for linear polymers \cite{Morseconstrainedtheory, ISI:000227372200078, ISI:A1995QD78900017, ISI:A1978FP21600029}, and generalized it to ring polymers. 
Our numerical results confirm earlier analytical results for flexible polymers and rigid rods. In both cases, though the tumbling dynamics is qualitatively very different, the tumbling frequency $f_c$ scales as $f_c \sim Wi^{2/3}$ in the Weissenberg number $Wi = \gamma \tau$. While for rigid rods the characteristic time $\tau$ scale is the global rotation time, it is the relaxation time of the longest Rouse mode for flexible coils. Our main finding is that for semiflexible polymers this simple scaling picture breaks down. The Weissenberg number no longer suffices to fully characterize the tumbling dynamics. In addition to the 
Weissenberg number $Wi = \gamma \tau^*$, where $\tau^*$ is the time a polymer needs to 
diffuse its own length, the tumbling frequency also depends on the polymer's relative stiffness $l_p/L$. Moreover, there is a distinct (intermediate) scaling law
\begin{equation}
f_c \tau^* = Wi^{3/4} {\hat f}_c (x)
\end{equation}
with the scaling variable $x = (l_p/L)Wi^{-1/3}$. The scaling function ${\hat f}_c$ exhibits a non-symmetrical peak until a crossover to Jeffery's theory is reached. Both of these features can be explained within a scaling picture analyzing the interplay between shear flow and bending modes which leads to two distinct types of buckling instabilities. 
Close to the Jeffery limit, there is a shear-induced global Euler buckling instability characterized by an ensuing overall bend conformation of the polymer. With decreasing polymer stiffness there is a second type of Euler instability where the quasi-stationary conformation of the polymer in shear flow exhibits a localized hairpin-like bend. These scaling pictures allowed us to rationalize the observed scaling regimes.

We have also compared our simulation results with recent experiments on single actin filaments in shear flow \cite{harasim2013direct}, and find quantitative agreement without any adjustable parameter. In particular, this shows that the $3/4$ scaling law fits the data significantly better than the classical $2/3$ scaling law. 
It would be interesting to test our numerical results also for microtubules and carbon nanotubes \cite{duggal2006dynamics, ericson2004macroscopic}. Actually, because of their length-dependent persistence length \cite{pampaloni2006thermal, taute2008microtubule, bathe2008cytoskeletal, heussinger2007} the dynamics of microtubules in shear flow may show an even richer scaling behavior than F-actin.

Finally, we have studied the tumbling dynamics of a ring polymer in shear flow. Similar to previous studies \cite{chen2013tumbling, chen2013effects} we find tow distinct types of tumbling events: rapid turnovers and tank-treading. The scaling behavior of the tumbling frequency is qualitatively similar to the results for linear polymers with slightly different power laws. 

It would be interesting to extend the theoretical analysis of semiflexible polymers beyond the scaling picture presented here and the force balance analysis given in Ref.~\cite{harasim2013direct}. We suppose, however, that this will pose significant technical challenges beyond the singular perturbation theory performed in Refs.~\cite{hallat, hallatschek2007tension, hallatschek2007tensionp2, obermayer2009tension, obermayer2009freely, thuroff2011longitudinal}  since the dynamics consists of two very distinct regimes: rapid tumbling events interrupted by extended quiescent periods in the neutral plane.

Further extension may account for a finite extensibility of the polymer backbone \cite{marko1998dna, kierfeld2004stretching, obermayer2009tension, netz2001strongly}, which one might expect to become important for very strong shear flow.

\begin{acknowledgments}
We acknowledge support by the Deutsche Forschungsgemeinschaft in the framework of the SFB 863 ``Forces in Biomolecular Systems''. We acknowledge fruitful discussions with Roland Winkler, Andreas Bausch and Bernhard Wunderlich.
\end{acknowledgments}

\cleardoublepage

\appendix

\section{Derivation of metric forces for a bead-rod algorithm of ring polymers}
\label{app:ringmetric}

In this appendix we derive the expression used for the metric forces required in the bead-rod algorithm to simulate ring polymers. For details on the metric force and a derivation of the metric force in general, see Refs.~\cite{Morseconstrainedtheory, ISI:000227372200078, effalgometric}. Here we follow closely the notation as introduced in  Ref.~\cite{effalgometric}, where the general expression for the metric force $\mathbf{F}^{\text{met}}_i$ on the $i$-th bead is given as
\begin{equation}
 \mathbf{F}^{\text{met}}_i = -\frac{1}{2} k_BT \frac{\partial \ln \det G}{\partial \mathbf{R}_i}, \label{def:Fmetric}
\end{equation}
with $\mathbf{R}_i$ the position of the $i$-th bead in a chain of $N$ beads.  The matrix $G$ represents the constraints of the system and is defined as \cite{Morseconstrainedtheory, ISI:000227372200078, effalgometric}
\begin{equation}
 G_{\mu \nu} = \sum_i \mathbf{n}_{i\mu} \mathbf{n}_{i\nu},
\end{equation}
where $\mathbf{n}_{i\mu} = \frac{\partial C_\mu}{\partial \mathbf{R}_i}$ is the derivative of the constraint $C_\mu$ with respect to the position of bead $i$. The constraints have to be of the form $C_\mu (\mathbf{R}_1,\dots,\mathbf{R}_N) = \text{const}$ for $\mu = 1,\dots, K$ with $K$ the number of constraints. For the polymers under consideration the constraints are of the form $C_\mu = |\mathbf{R}_\mu - \mathbf{R}_{\mu+1}| =b$ such that the distance between two beads is equal to the bond length $b$.
For a linear chain $\mu = 1,\dots, N-1$, and it follows that the matrix $G$ is a tridiagonal, symmetric matrix of the form 
\begin{equation}
 G= \left( \begin{array}{cccccccc}
	    d & c_2 & 0 & \cdots & \ddots &  & & \\
	    c_2 & d & c_3 & 0 & \ddots & && \\
	    0& c_3 & d & c_4 & 0 & \ddots& &\\
	     \vdots &\ddots & \ddots & \ddots & \ddots && \\
	    & &0 & c_{N-3} & d & c_{N-2} &0 \\
	    && &0 & c_{N-2} & d & c_{N-1} \\
	    && \ddots && 0 & c_{N-1} & d  \\
        \end{array}
\right),
\end{equation}
where $d=2$ and $c_i=-\mathbf{u}_i \cdot \mathbf{u}_{i-1}$ for $i=2,\dots N-1$ \cite{effalgometric}.  Here $\mathbf{u}_i=(\mathbf{R}_{i+1}-\mathbf{R}_{i})/b$ for $i=1,\dots N-1$ is the normalized bond vector. Due to this specific form the metric force may be recast to \cite{effalgometric}
\begin{equation}
 \mathbf{F}^{\text{met}}_k = k_BT \sum_{i=2}^{N-1}  G_{i-1,i}^{-1} \frac{\partial (\mathbf{u}_i \cdot \mathbf{u}_{i-1})}{\partial \mathbf{R}_k}, \label{eq:Flin}
\end{equation}
where $G_{i-1,i}^{-1}$ is the $(i-1,i)$ component of the inverse Matrix of $G$.
This has the advantage that it may be evaluated by an efficient algorithm linear in the number of beads \cite{effalgometric}.

Here, we aim at keeping this advantage for the algorithm to simulate ring polymers. For a ring there are $N$ constraints on the bond length, which are of identical form as the constraints for linear polymers. We use periodic boundary conditions by setting $\mathbf{R}_N+1 = \mathbf{R}_1$ such that we can keep the previously stated form of $C_\mu$, now with $\mu= 1,\dots, N$. Analogously the $N$-th bond vector is defined $\mathbf{u}_N=(\mathbf{R}_{N+1}-\mathbf{R}_{N})/b =(\mathbf{R}_{1}-\mathbf{R}_{N})/b$.
Hence the matrix $G$ is now of rank $N \times N$ and symmetric. However, in contrast to the linear case $G$ is cyclic instead of tridiagonal for a ring polymer.
With these definitions, one obtains 
\begin{equation}
G= \left( \begin{array}{ccccccccc}
	    d & c_2 & 0 & \cdots & \ddots &  & & 0 & c_1 \\
	    c_2 & d & c_3 & 0 & \ddots & && &0 \\
	    0& c_3 & d & c_4 & 0 & \ddots& & &\\
	     \vdots &\ddots & \ddots & \ddots & \ddots &&  &\\
	    & & &0 & c_{N-3} & d & c_{N-2} &0  \\
	    & && &0 & c_{N-2} & d & c_{N-1}&0 \\
	    0& && \ddots && 0 & c_{N-1} & d &c_N \\
	    c_1&0 & && \ddots && 0 & c_{N} & d  \\
        \end{array}
\right),
\end{equation}
where analogously to the linear chain $d=2$ and $c_i=-\mathbf{u}_i \cdot \mathbf{u}_{i-1}$ for $i=2,\dots, N$, and we additionally define $c_1=-\mathbf{u}_1 \cdot \mathbf{u}_{N}$ in accord with the periodic boundary conditions.

Starting from the general expression, Eq.~(\ref{def:Fmetric}), we analogously to the linear case may reformulate the metric force to 
\begin{equation}
 \mathbf{F}^{\text{met}}_k = k_BT \sum_{i=2}^{N}  G_{i-1,i}^{-1} \frac{\partial (\mathbf{u}_i \cdot \mathbf{u}_{i-1})}{\partial \mathbf{R}_k} + k_BT G_{N,1}^{-1} \frac{\partial (\mathbf{u}_1 \cdot \mathbf{u}_{N})}{\partial \mathbf{R}_k} .
\end{equation}
As compared to to Eq.~(\ref{eq:Flin}) for linear polymers there is an additional term ensuing from the additional constraint. Since it is of the same mathematical structure as the first term we may use an adjusted form of the algorithm developed in Ref.~\cite{effalgometric}. The basic idea of this algorithm is to avoid the $\mathcal{O}(N^3)$ inversion of $G$ by only evaluating the required entries of $G^{-1}$. This is achieved by using Cramer's rule and expressing the components of the inverse Matrix by the determinant of $G$ and the determinant of the matrices resulting from removing a row and a column from $G$.
For this task a linear, iterative scheme is developed in Ref.~\cite{effalgometric}, which is based on the special property of $G$ being tridiagonal. 

To adjust this algorithm for ring polymers with a cyclic matrix $G$ we use Cramer's rule to transform the cyclic matrix to a tridiagonal or otherwise trivial matrices.
Hence we get  
\begin{widetext}
\begin{eqnarray*}
\det G &=& \\
&-&c_{1}^2 \det \left( \begin{array}{cccccccc}
	    d & c_3 & 0 & \cdots & \ddots &  & &    \\
	    c_3 & d & c_4 & 0 & \ddots & &&  \\
	    0& c_4 & d & c_5 & 0 & \ddots& & \\
	     \vdots &\ddots & \ddots & \ddots & \ddots &&  \\
	     & &0 & c_{N-4} & d & c_{N-3} &0  \\
	     && &0 & c_{N-3} & d & c_{N-2}&0 \\
	     && \ddots && 0 & c_{N-2} & d &c_{N-1} \\
	     & && \ddots && 0 & c_{N-1} & d  \\
        \end{array}
\right)
- c_N^2  \det \left( \begin{array}{cccccccc}
	    d & c_2 & 0 & \cdots & \ddots &  & &    \\
	    c_2 & d & c_3 & 0 & \ddots & &&  \\
	    0& c_3 & d & c_4 & 0 & \ddots& & \\
	     \vdots &\ddots & \ddots & \ddots & \ddots &&  \\
	     & &0 & c_{N-5} & d & c_{N-4} &0  \\
	     && &0 & c_{N-4} & d & c_{N-3}&0 \\
	     && \ddots && 0 & c_{N-3} & d &c_{N-2} \\
	     & && \ddots && 0 & c_{N-2} & d  \\
        \end{array}
\right) \\
&+&d \det \left( \begin{array}{ccccccccc}
	    d & c_2 & 0 & \cdots & \ddots &  & &  &  \\
	    c_2 & d & c_3 & 0 & \ddots & && &0 \\
	    0& c_3 & d & c_4 & 0 & \ddots& & &\\
	     \vdots &\ddots & \ddots & \ddots & \ddots &&  &\\
	    & & &0 & c_{N-4} & d & c_{N-3} &0  \\
	    & && &0 & c_{N-3} & d & c_{N-2}&0 \\
	    & && \ddots && 0 & c_{N-2} & d &c_{N-1} \\
	    & & && \ddots && 0 & c_{N-1} & d  \\
        \end{array}
\right) \, .
\end{eqnarray*}
\end{widetext}
With this form we may use the iterative scheme from the linear chain to calculate the determinant of $G$ in the ring case, also.

Analogously we transform the matrices after removing row $i$ and column $i-1$ as required for the inverse of $G$ at $(i-1,i)$. We get
\begin{widetext} 
\begin{eqnarray*}
 \det \left( \begin{array}{ccccccccc}
	    d & c_2 & 0 & \cdots & \ddots &  & & 0 & c_1 \\
	    c_2 & d & c_3 & 0 & \ddots & && &0 \\
	     \ddots &\ddots & \ddots &  &  &&  &\\
	    &0& c_{i-1} & d & 0 & 0 & \ddots& & \\
	     & &0 & c_{i} & c_{i+1} &0 & &  \\
	     && &0 & c_{i+2} & d & c_{i+3}&0& \\
	    &  &  &&  & \ddots &\ddots & \ddots \\
	    0& && \ddots && 0 & c_{N-1} & d &c_N \\
	    c_1&0 & && \ddots && 0 & c_{N} & d  \\
        \end{array}
\right) =& \\
-c_{1}^2 \det \left( \begin{array}{ccccccccc}
	    d & c_3 & 0 & \cdots & \ddots &  & & &   \\
	    c_3 & d & c_4 & 0 & \ddots & && & \\
	     \ddots &\ddots & \ddots &  &  &&  &\\
	    &0& c_{i-1} & d & 0 & 0 & \ddots& & \\
	     & &0 & c_{i} & c_{i+1} &0 & &  \\
	     && &0 & c_{i+2} & d & c_{i+3}&0& \\
	    &  &  &&  & \ddots &\ddots & \ddots \\
	    & && \ddots && 0 & c_{N-2} & d &c_{N-1} \\
	    & & && \ddots && 0 & c_{N-1} & d  \\
        \end{array}
\right)&
- c_N^2  \det \left( \begin{array}{ccccccccc}
	    d & c_2 & 0 & \cdots & \ddots &  & & &   \\
	    c_2 & d & c_3 & 0 & \ddots & && & \\
	     \ddots &\ddots & \ddots &  &  &&  &\\
	    &0& c_{i-1} & d & 0 & 0 & \ddots& & \\
	     & &0 & c_{i} & c_{i+1} &0 & &  \\
	     && &0 & c_{i+2} & d & c_{i+3}&0& \\
	    &  &  &&  & \ddots &\ddots & \ddots \\
	     &&& \ddots && 0 & c_{N-3} & d &c_{N-2} \\
	     && && \ddots && 0 & c_{N-2} & d  \\
        \end{array}
\right) \\
+d \det \left( \begin{array}{ccccccccc}
	    d & c_2 & 0 & \cdots & \ddots &  & &  &  \\
	    c_2 & d & c_3 & 0 & \ddots & && &0 \\
	     \ddots &\ddots & \ddots &  &  &&  &\\
	    &0& c_{i-1} & d & 0 & 0 & \ddots& & \\
	     & &0 & c_{i} & c_{i+1} &0 & &  \\
	     && &0 & c_{i+2} & d & c_{i+3}&0& \\
	    &  &  &&  & \ddots &\ddots & \ddots \\
	    & && \ddots && 0 & c_{N-2} & d &c_{N-1} \\
	    & & && \ddots && 0 & c_{N-1} & d  \\
        \end{array}
\right)& +  \prod_{1\le k \le N \atop k \ne i}\limits c_k \, .
\end{eqnarray*}
\end{widetext}
These two expressions for all required determinants are tridiagonal and hence compatible with the algorithm of Ref.~\cite{effalgometric} or trivial to calculate. Hence we are able to calculate all required values using two times the iterative scheme introduced there, and get an efficient algorithm which is $\mathcal{O}(N)$.

\section{Parameters used in simulations}
Here we give a complete list of the parameters used in our simulations for generating the results as presented in the paper. As stated there, we set $k_BT = \zeta =1$ for all simulations. Each chain consists of $N$ beads with bond-length $\delta r$ and is simulated with a time-step $\delta t$. For the relaxation dynamics of ring polymers we used the following set of parameters:
\begin{center}
 \begin{tabular}{c|c|c|c}
 \begin{minipage}{2cm}
  used in Fig.
 \end{minipage}
& $N$ & $\delta r$ & $\delta t$ \\
\hline
1 & 10 & 1 & 0.0001 \\
1 & 20 & 1 & 0.0000001 \\
1 & 50 & 1 & 0.00000125 \\
\hline
2 & 20 & 1 & 0.0000001 
\end{tabular}
\end{center}
The linear relaxation curve in Fig.~2 was generated using $N=21$, $\delta r=0.3$, $\delta t=0.000001$.

For the simulation of the tumbling in shear flow, the strength of the shear may be determined from the dimensionless shear rate in the paper using the parameters of the polymers as given here. We used the following set of parameters: \\
\begin{center}
 \begin{tabular}{c|c|c|c | c}
 \begin{minipage}{2cm}
  used in Fig.
 \end{minipage}
& $N$ & $\delta r$ & $\delta t$ & $l_p$ \\
\hline
4 & 100 & 0.3 & 0.000001 & 60 \\
\hline
5 & 100 & 0.25 & 0.00002 & 2\\
5 & 200 & 0.25 & 0.0000125 & 4\\
5 & 100 & 0.75 & 0.00002 & 6\\
5 & 50 & 0.5 & 0.000005 & 63\\
5 & 50 & 1 & 0.00004 & 125\\
5 & 100 & 0.75 & 0.00003 & 188\\
\end{tabular}
\end{center}
When varying the persistence length we adjusted the parameters in the regions of extremal values of stiffness due to numeric stability. The data was generated using the values
\begin{center}
 \begin{tabular}{c|c|c|c|c }
 \begin{minipage}{2cm}
  used in Fig.
 \end{minipage}
& $N$ & $\delta r$ & $\delta t$ & $l_p$ \\
\hline
6/7 & 17 & 1.5 & 0.00001 & 1000-30000\\
6/7 & 34 & 0.75 & 0.00005 & 15-5000\\
6/7 & 102 & 0.25 & 0.00001 & 1-30\\
6/7 & 255 & 0.1 & 0.000007 & 0.1-10\\
\hline
8 & 120 & 0.25 & 0.0000067 & 1.2 \\
8 & 60 & 0.5 & 0.000002 & 6 \\
8/9 & 60 & 0.5 & 0.000004 & 18 \\
8 & 30 & 1 & 0.000005 & 60 \\
\hline
8 & 20 & 1.5 & 0.000004 & 5000 \\
8 & 20 & 1.5 & 0.000002 & 50000 \\
\end{tabular}
\end{center}
For the simulation of ring polymers under shear flow we used:

\begin{center}
 \begin{tabular}{c|c|c|c|c }
 \begin{minipage}{2cm}
  used in Fig.
 \end{minipage}
& $N$ & $\delta r$ & $\delta t$ \\
\hline
10 & 80 & 1 & 0.0000125 \\
\hline
11/12 & 80 & 1 & 0.000005 \\
\end{tabular}
\end{center}

\section{Calculation of the internal relaxation of a semiflexible ring polymer}
\label{app:ringanalytisch}

To determine the contribution of internal relaxation to the mean square deviation of the diameter for a semiflexible ring polymer $\delta \mathbf{D}^2_{\mathrm{shape}}$, we apply the weakly bending limit and follow closely the calculations as known for linear polymers \cite{frey1991dynamics, kroy1997dynamic, ISI:A1996TN90500006, farge1993dynamic, granek1997semi, hallat} and outlined in Sec.~\ref{sec:EqDynamics}. 
In thermal equilibrium the contour of a semiflexible ring polymer with $l_p >L$ is effectively constrained to a plane and acquires a shape deviating only slightly from a circle. Therefore, we use cylindrical coordinates and choose the origin such that the polymer ring is in the $xy$-plane. Analogously to the Monge parameterization for linear polymers \cite{frey1991dynamics, kroy1997dynamic, ISI:A1996TN90500006, farge1993dynamic, granek1997semi, hallat} we introduce the following parameterization for the ring polymer: 
\begin{equation}
 \mathbf{r}(s)= \left(R+ \rho(s) \right) \hat{\mathbf{e}}_r(s) + \phi(s) \hat{\mathbf{e}}_\phi(s) + z(s) \hat{\mathbf{e}}_z(s) \ ,
\end{equation}
where $R=L/2 \pi$ and $\hat{\mathbf{e}}_r$, $\hat{\mathbf{e}}_\phi$ and $\hat{\mathbf{e}}_z$ are the unit vectors in $r$, $\phi$ and $z$ direction, respectively. The variables $\rho(s)$, $\phi(s)$ and $z(s)$ characterize the small perturbation relative to the idealized circle-like conformation. As for linear polymers these perturbations are coupled by the inextensiblity constraint. Analogously to the relation for longitudinal and transversal fluctuations in the linear case one can show that the perturbations in $\rho(s)$ are dominant for $\delta \mathbf{D}^2_{\mathrm{shape}}$, whereas $\phi$ and $z$ are of higher order. 
Using the parameterization and separating the equation of motions for the perturbations analogous to the linear case in Eq.~\ref{eq:langevinrperplinear}, we arrive at
\begin{equation}
 \zeta \frac{\partial \mathbf{\rho}(s,t)}{\partial t} = - \kappa \left( \frac{\partial^4 \, \mathbf{\rho}(s,t)}{\partial s^4} + \frac{1}{R^4}\rho  + \frac{2}{R^3} \right) + \eta_\rho (s,t) \ ,
\end{equation}
where for the noise $\eta_\rho (s,t)$ it holds $\langle \eta_\rho (s,t)  \rangle =0$ and $\langle \eta_\rho (s,t) \, \eta_\rho (s',t') \rangle = 2  \, \zeta \, k_B T \, \delta(s-s') \, \delta(t-t').$ 
We solved this equation by a linear mode analysis using the modes $\sin \left( sn/R \right)$ and  $\cos \left( sn/R \right)$, where $n$ is the mode number, to obtain Eq.~\ref{eq:msd-shape}.

\cleardoublepage

\end{document}